\documentstyle[11pt,newpasp,twoside,epsf]{article}
\def\edcomment#1{\iffalse\marginpar{\raggedright\sl#1\/}\else\relax\fi}
\reversemarginpar

\newcommand{\teff}  {\mbox{$T_{\rm eff}$}}

\newcommand{\logg}  {\mbox{{\rm log}\,$g$}}

\begin{document}
\title{Uncertainties in stellar abundance analyses}
 \author{Martin Asplund}
\affil{Research School of Astronomy and Astrophysics,
       Mt Stromlo Observatory, Cotter Rd, ACT 2611, Australia 
      (martin@mso.anu.edu.au)}

\begin{abstract}

Over the last half-century quantitative stellar spectroscopy
has made great progress. However, most stellar abundance analyses
today still employ rather simplified models, which can introduce
severe systematic errors swamping the observational errors. 
Some of these uncertainties for late-type stars are briefly reviewed here:
atomic and molecular data, stellar parameters,
model atmospheres and spectral line formation.

\end{abstract}

\section{Introduction}

In view of the central role stellar abundance analyses play 
in the endeavours to decipher the formation and evolution of
stars, galaxies and indeed the Universe as a whole, minimizing
systematic errors 
should be of utmost importance. Certainly, there are 
many potential fallacies that can be made in the process
of going from an observed stellar spectrum to the extracted
chemical composition of the star, all which deserves 
very careful consideration.
Unfortunately, this is an area which often has not received
the attention its importance warrants.
Instead, still today most elemental abundance analyses  
of late-type stars rely on very simplified 
models for the stellar atmospheres and the spectral
line formation processes.
Unfortunately, the progress in modelling has not kept
up with the dramatic improvements on the observational side
over the last couple of decades, leaving the error budget
normally dominated by systematic uncertainties. 

Due to page restrictions this  review focus only on the
uncertainties in the derived elemental abundances introduced
during the numerical analyses.
Potential observational pitfalls such as signal-to-noise,
resolving power, fringing, scattered light, continuum placement and blends can
certainly also be major sources of error, but
are not discussed here. 
Furthermore, the review is limited to late-type stars
as they have traditionally been the most widely used beacons when
tracing Galactic chemical evolution. The reader is referred to
Werner et al. (2002) for an account of current hot star modelling,
which is becoming increasingly important when probing environments beyond
our own Galaxy.

\section{Atomic and molecular data}

The most obvious input data needed to derive elemental abundances
is the transition probability, normally expressed as the $gf$-value.
While there is always a continuing need for more and better
data in this respect, the {\em overall situation} is in fact 
relatively good today. 
Provided the stellar spectroscopists are prepared to search the
physics literature and databases, there are many accurate experimental and
computational $gf$-determinations available
(e.g. http://physics.nist.gov).
The Kurucz database (http://kurucz.harvard.edu)
is a very valuable resource but the drawback with such a large-scale
computational effort is that individual transitions can be very erroneous,
in particular when involving predicted energy levels.  
Other necessary data (continuous opacities, line broadening,
dissociation energies etc) are also {\em in general} in reasonably healthy
shape now (e.g. Seaton et al. 1994; Barklem et al. 2000), 
although improvements are certainly encouraged.  

As will be discussed further below, the formation of a spectral line
depends in principle not only on the line itself 
but on all other lines also, including those of other elements.
In order to compute the statistical equilibrium of a species one
needs not only transition probabilities for all relevant lines but
also photo-ionization and collisional cross-sections. 
In terms of photo-ionization there has been marked improvements
recently with the advent of large opacity calculations like
Opacity Project and Iron Project (e.g. Seaton et al. 1994).
For elements up to the Fe-peak the situation is now fairly
healthy for late-type stars. The most pressing uncertainty 
in non-LTE studies today is the cross-sections for collisional
excitation and ionization with electrons and hydrogen atoms. 
The Opacity Project has partly addressed the case of
electron collisions but most calculations largely rely on classical 
recipes like van Regemorter's (1963) formula. 
The situation for inelastic H collisions is even worse
with the approach of Drawin (1968) mostly used. 
The few existing experimental and quantum mechanical calculations
suggests, however,  that the Drawin recipe over-estimates
the cross-sections by about three orders of magnitude,
at least for Na and Li (e.g. Fleck et al. 1991; Barklem et al. 2003). 
Whether this is true
for all elements is not known. Clearly there is a great
need for more quantum mechanical calculations addressing
this fundamental problem.

\section{Stellar parameters}

We will here limit the discussion to methods more universally used, noting that
in special situations other more accurate options are available
(interferometry, eclipsing binaries etc).

Of the fundamental stellar parameters, $T_{\rm eff}$ is normally
the most crucial in order to obtain accurate abundances. There exists a multitude
of methods to determine $T_{\rm eff}$ of varying model dependence and reliability.
Of these, the infrared flux method 
(IRFM, Blackwell \& Shallis 1977)
is often advocated as the best.
IRFM is based on the ratio of bolometric flux ($\propto T_{\rm eff}^4$ and reddening-
and model-independent) with an IR monochromatic flux ($\propto T_{\rm eff}$ and
essentially reddening- and model-independent). If the problem of collecting
sufficiently accurate (spectro-)photometry can be overcome, IRFM should yield
temperatures to better than 50\,K (Alonso et al. 1996).
Photometric $T_{\rm eff}$ determinations can be almost as good when using
colours like $V-K$ and $b-y$ (corrected for interstellar reddening
if significant), in particular
if calibrated to an IRFM- or interferometric temperature scale
(Bessell et al. 1998).
As always with theoretical colours, the zero-point is an outstanding issue.

In principle, hydrogen Balmer lines should be sensitive thermometers but
practical problems unfortunately limit their usefulness, not the least
observational. The Balmer lines
are formed in deep atmospheric layers where convective energy transport is
important for setting the temperature structure. The classical mixing length
theory for convection in 1D models is unlikely to capture all aspects in
this transition from convection to radiation, as will be further discussed below.
The line broadening, including self-broadening, of the H lines have recently
been improved (Barklem et al. 2002) but these results have not yet
been fully disseminated into the wider astronomical community, leading to
unnecessary additional errors. At solar metallicities H lines could
in the best cases yield $T_{\rm eff}$ to within about 100\,K but the
uncertainties become progressively worse for metal-poor stars.
The use of for example excitation balance of Fe\,{\sc i} and other lines
and various line-depth ratios can achieve highly precise {\em relative}
temperatures for similar stars but due to possible non-LTE and 3D effects
can not be expected to give accurate {\em absolute} values.
In summary, $T_{\rm eff}$ can under favourable circumstances be determined
to within 100\,K, which corresponds to abundance errors of typically
0.1\,dex.

The surface gravities are often the most poorly constrained
parameter. With the advent of the Hipparcos astrometry the situation has
improved dramatically, at least for the stars sufficiently nearby to show
measurable parallaxes.
Knowing the parallax, the observed magnitude can be
converted to a surface gravity:
$ {\rm log} g / g_{\sun}  =  {\rm log} {\cal M} / {\cal M}_{\sun}
 + 4 \cdot {\rm log} \teff / T_{{\rm eff,}\sun}
 + 0.4 \cdot (M_{\rm bol} - M_{{\rm bol,}\sun}).$
The uncertainty is normally dominated by the parallax error 
going into $M_{\rm bol}$ but if
$\Delta \pi/\pi < 0.2$, log\,$g$ can be determined to within 0.2\,dex
(e.g. Nissen et al. 1997).
The Str\"omgren $c_1$-index and isochrone-fitting can also be employed
to estimate log\,$g$ but are more uncertain.
The pressure-sensitive wings of strong lines, such as the Mg\,{\sc i}b triplet
(Blackwell \& Willis 1977) is a good gravity-meter with the caveat
of potential systematic errors due to non-LTE and 3D effects, which
have not yet been fully assessed.  
One of the most commonly used techniques is to force ionization balance
between neutral and ionized species such as Fe\,{\sc i}/Fe\,{\sc ii}
and Ti\,{\sc i}/Ti\,{\sc ii}.
In the absence of any of the above-mentioned procedures, this is a viable option
but it must be realised that the result may be very severe errors in log\,$g$.
Due to over-ionization of neutral minority species (e.g. Fe\,{\sc i})
compared with the LTE predictions, this method typically underestimates
the gravity by up to 0.5\,dex or even more (Thevenin \& Idiart 1999).
For pressure-sensitive spectral
features like molecular lines this can obviously be disastrous.
A good habit is to ratio two equally gravity-sensitive species like
C\,{\sc i}, O\,{\sc i}, S\,{\sc i} and Fe\,{\sc ii}
to obtain abundance ratios (e.g. Nissen et al. 2002, 2003; Akerman et al. 2003).

In view of the potentially large non-LTE effects and 3D effects on
Fe\,{\sc i} lines discussed below, the preferred choice for the
metallicity determinations is no doubt Fe\,{\sc ii} lines, which are largely immune
to such problems (e.g. Thevenin \& Idiart 1999; Asplund et al. 1999).
This should yield [Fe/H] values accurate to typically within 0.1-0.2\,dex,
depending on how well $T_{\rm eff}$ and log\,$g$ can be constrained.
The alternative method on relying on colours, in particular Str\"omgren
photometry, normally gives reasonable results with uncertainties $\la 0.3$\,dex
when properly calibrated.
Any error in [Fe/H] naturally directly propagates into the derived
[X/Fe] ratios, re-enforcing the need for a simultaneous [Fe/H] determination
together with the other elements rather than relying on literature values.

\section{Stellar model atmospheres}

The vast majority of abundance analyses of late-type stars rely
on model atmospheres which are 1D, time-independent and hydrostatic,
which assume LTE and treat convection with the
rudimentary mixing length theory. Even a casual glance at the solar
surface reveals that these assumptions and approximations are
very disputable. The question is whether this propagates into
significant systematic errors in the derived abundances?
Recently realistic 3D time-dependent hydrodynamical simulations of
stellar surface convection and atmosphere with a detailed treatment
of radiative transfer and state-of-the-art equation-of-state and
opacities have become available for solar-type stars
(e.g. Nordlund \& Dravins 1990; Stein \& Nordlund 1998;
Asplund et al. 1999; Asplund \& Garc\'{\i}a P{\'e}rez 2001). 
They successfully reproduce a wide range of observational diagnostics
(granulation topology, helioseismology, intensity brightness contrasts,
spectral line shapes, shifts and asymmetries etc). 
It therefore appears that one can place a fairly high degree of confidence 
in their ability in describing the real stellar atmospheres, in spite
of the simplifications necessary in order to carry out the
simulations, most notably in terms of numerical resolution and
radiative transfer.
A notable achievement is that the traditional free parameters
of stellar spectroscopy 
(mixing length parameters, micro- and macroturbulence)
have become obsolete with 3D models, greatly 
reducing the uncertainties.

This new generation of 3D hydrodynamical model atmospheres
have started to be applied to stellar abundance analyses.
For the Sun, this has caused a very substantial reduction 
(0.2-0.3\,dex) in the solar C, N and O abundances
(Allende Prieto et al. 2001, 2002; Asplund et al. 2003b).
For the first time, all different diagnostics (permitted,
forbidden and molecular lines) give concordant abundances.
The new results are also supported by the excellent
agreement between observed and predicted line shapes
and center-to-limb variations. 
Other, less temperature-sensitive elements like Si and Fe
only show small ($\la 0.05$\,dex) 3D abundance corrections 
(Asplund et al. 2000).
The exact 3D effects depend on the ionization stage, excitation
potential and strength of the line in question. 

The most dramatic differences with standard 1D analyses 
appears at low metallicities. Due to much lower temperatures
in the optically thin layers in the metal-poor 3D models as a result
of the shift in balance between expansion cooling and radiative
heating, many spectral features are greatly affected 
(Asplund et al. 1999). 
In particular molecular lines but
also low excitation lines and neutral minority species
tend to have large negative 3D abundance corrections 
in metal-poor stars (i.e. 1D analyses {\em over-estimate} 
the abundances). As a result, Fe\,{\sc i} lines are very unreliable
but Fe\,{\sc ii} lines, which are formed in deeper atmospheric
layers where the differences between 3D and 1D models are much 
smaller, are quite robust. 
In some cases additional 3D non-LTE effects can 
conspire to give final results quite close to 
the 1D non-LTE case, as for Li (Asplund et al. 2003a),
but this is obviously not generally true. In the absence 
of detailed 3D non-LTE calculations, we advise against
using Fe\,{\sc i} and such species. Extreme caution
must be exercised when relying on resonance and other
low excitation lines (Al, Mg, Sr, Ba, Eu etc) in halo stars,
where the systematic errors may well be $-0.3$..$-0.5$\,dex. 
The largest errors, however, occur for molecular lines
(Fig. 1)
for which the 1D analyses can overestimate the abundances
by up to 1.0\,dex at [Fe/H]\,$=-3$
(Asplund \& Garc\'{\i}a P{\'e}rez 2001; Asplund 2003).
Needless to say, such large systematic errors can have a
profound impact on the interpretations in terms of
stellar nucleosynthesis and galactic chemical evolution. 

\begin{figure}
\plotfiddle{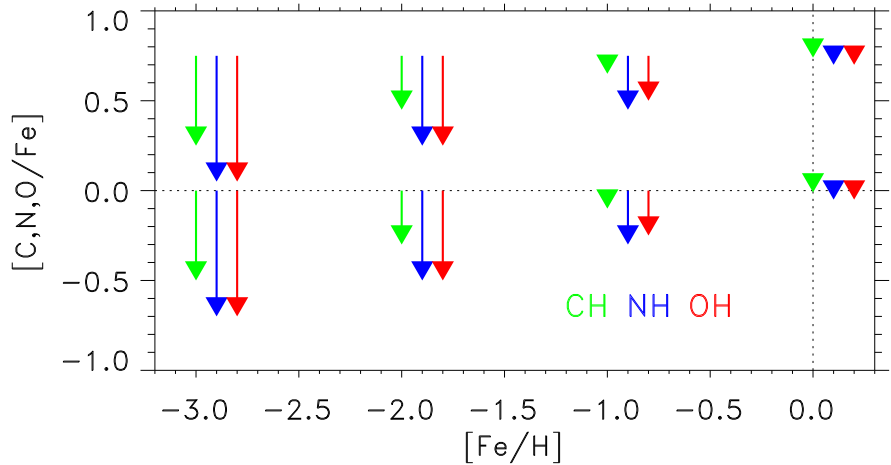}{7cm}{0}{70}{70}{-180}{0}
\caption{The typical 3D abundance corrections {\em relative to the Sun}
for CH, NH and OH (A-X) lines reveal a strong metallicity dependence.
\label{f:CNO} }
\end{figure}

\section{Spectral line formation}

Spectral line formation essentially always occurs
as a non-equilibrium process:
under typical atmospheric conditions radiative rates
dominate over collisional rates and the radiation
field departs from the Planck function.
{\em Non-LTE line formation} is therefore neither
special nor unusual, while {\em LTE line formation} is:
LTE is an extreme assumption, not a cautious middle-ground.
Of course, in many incidences the different line
formation processes are such that LTE-based abundances 
are indeed good approximations
 but that must always be confirmed a posteriori
by detailed non-LTE calculations (e.g. Fe\,{\sc ii}). 
In general, non-LTE effects become progressively worse for
higher \teff\ (higher $J_\nu$) and lower \logg\ (less collisions)
and [Fe/H] (less e$^-$ collisions and stronger UV radiation field).

In spite of the availability of efficient and user-friendly
non-LTE codes such as {\sc multi} (Carlsson 1986), not enough work 
has been devoted to this important area. 
Amazingly, detailed non-LTE studies of solar-type and metal-poor stars
have been undertaken for only a dozen elements or so. 
Typical non-LTE abundance corrections for halo stars are 
$0.2-0.3$\,dex of either
sign (e.g. Be\,{\sc ii}, O\,{\sc i}, Mg\,{\sc i}, K\,{\sc i}, Ca\,{\sc i},
Fe\,{\sc i}, Sr\,{\sc ii}, Ba\,{\sc ii}) but 
significantly larger in cases like Al\,{\sc i} and B\,{\sc i} 
(Kiselman 1994).
It is true that the poorly known H collision cross-sections
introduce uncertainties
(e.g. Korn et al. 2003) 
but calculations with and without the classical Drawin (1968)
recipe should bracket the expected non-LTE effects; 
as already mentioned, the available evidence suggests that
the Drawin formula over-estimates the H collisions by
about three orders of magnitude.

The lack of non-LTE calculations for the majority of elements
severely hampers our understanding of stellar nucleosynthesis
and galactic chemical evolution. 
For example, it is not known whether the recently discovered  
upturn in [C/O] at the lowest [Fe/H] is due to C production 
in Pop III stars or can be explained by differential non-LTE effects
between C\,{\sc i} and O\,{\sc i} (Akerman et al. 2003).
Clearly, there is huge need for more non-LTE investigations
for more elements. 


\acknowledgments

I thank the organizers (Poul Erik Nissen and Max Pettini) and
the other JD15 participants for creating a stimulating meeting.
This talk is available at 
http://www.mso.anu.edu.au/$\sim$martin/talks/Sydney03\_jd15).

\end{document}